\documentclass[10pt,a4paper,conference]{IEEEtran}

\usepackage{graphicx}
\usepackage{epstopdf}
 \usepackage{verbatim}
\usepackage{times}
\usepackage{amsmath}
\usepackage{amsfonts}
\usepackage{subfigure}
\usepackage{mathrsfs}
\usepackage{url}
\usepackage[]{amsmath}
\usepackage[numbers,sort&compress]{natbib}

\begin{document}
\newcounter{mytempeqncnt}
\title{Directional Modulation Design Based on Crossed-Dipole Arrays for Two Signals With Orthogonal Polarisations}
\author{\IEEEauthorblockN{Bo Zhang, Wei Liu$^1$ and Xiang Lan}
\IEEEauthorblockA{Communications Research Group, Department of Electronic and Electrical Engineering\\University of Sheffield, Sheffield S1 4ET, United Kingdom\\
$^1$ w.liu@sheffield.ac.uk}}

\maketitle
\begin{abstract}
Directional modulation (DM) is a physical layer security technique based on antenna arrays and so far the polarisation information has not been considered in its designs. To increase the channel capacity, we consider exploiting the polarisation information and send two different signals simultaneously at the same direction, same frequency, but with different polarisations. These two signals can also be considered as one composite signal using the four dimensional (4-D) modulation scheme across the two polarisation diversity channels. In this paper, based on cross-dipole arrays, we formulate the design to find a set of common weight coefficients to achieve directional modulation for such a composite signal and examples are provided to verify the effectiveness of the proposed method.

\end{abstract}



%
\IEEEpeerreviewmaketitle

\section{Introduction}
Directional modulation (DM) is a physical layer security technique which keeps known constellation mappings in a desired direction or directions, while scrambling them for the remaining ones~\cite{babakhani08b,babakhani09a}. Both reconfigurable arrays and phased array structures can be used for its design. For reconfigurable arrays, their elements are switched for each symbol to make their constellation points not scrambled in desired directions, but distorted in other directions~\cite{daly10a}. For phased arrays~\cite{daly09a,zhang17a}, DM can be implemented by phase shifting the transmitted antenna signals properly. In~\cite{hong11a}, a method named dual beam DM was introduced, where the I and Q signals are transmitted by different antennas. The bit error rate (BER) performance of a system based on a two-antenna array was studied using the DM technique for eight phase shift keying modulation in~\cite{shi13a}. A more systematic pattern synthesis approach was presented in~\cite{ding13b}, followed by an orthogonal vector approach which produces an artificial orthogonal noise vector from information to achieve DM in~\cite{ding14b}, and energy-constrained design in~\cite{ding14a}. Recently, the time modulation technique was introduced to DM to form a four-dimensional (4-D) antenna array in~\cite{zhu14a}.

However, the polarisation information is not exploited in all aforementioned DM designs, and a single signal is transmitted through the propagation channel. To increase system capacity, we can exploit the polarisation information of the electromagnetic signal and send two different signals simultaneously at the same direction, same frequency, but different polarisations. This can be achieved by employing polarisation-sensitive arrays, such as tripole arrays and crossed-dipole arrays~\cite{compton81a,nehorai98a,miron06a,gou11a,liu14a,hawes15a,xiao09a}. These two signals can also be considered as one composite signal using the four dimensional (4-D) modulation scheme across the two polarisation diversity channels~\cite{liu14a,isaeva95a,zetterberg77a}. In this paper, based on this idea, we proposed a DM design using crossed-dipole arrays, which can send two separate signals ($S_1$ and $S_2$) with orthogonal polarisation states to the same direction simultaneously. To receive and separate the two orthogonally polarised signals, at the receiver side, a crossed-dipole antenna or array is needed, similar to the transmitter side~\cite{lan17a}. However, the polarisation directions of the antennas at the receiver side do not need to match those of the transmitted signals, as cross-interference due to a mismatch can be solved easily using some standard signal processing techniques~\cite{liu14a}.

The remaining part of this paper is structured as follows. A review of polarised beamforming based on crossed-dipole arrays is given in Sec. \ref{sec:Review of Polarised Beamforming}. Directional modulation designs for two signals transmitted to the same direction, but with orthogonal polarisation states are presented in Sec. \ref{sec:Directional Modulation Design}. In Sec. \ref{sec:sim}, design examples are provided, with conclusions drawn in Sec. \ref{sec:con}.

\section{Review of Polarised Beamforming}\label{sec:Review of Polarised Beamforming}

\begin{figure}
  \centering
  \includegraphics[width=0.45\textwidth]{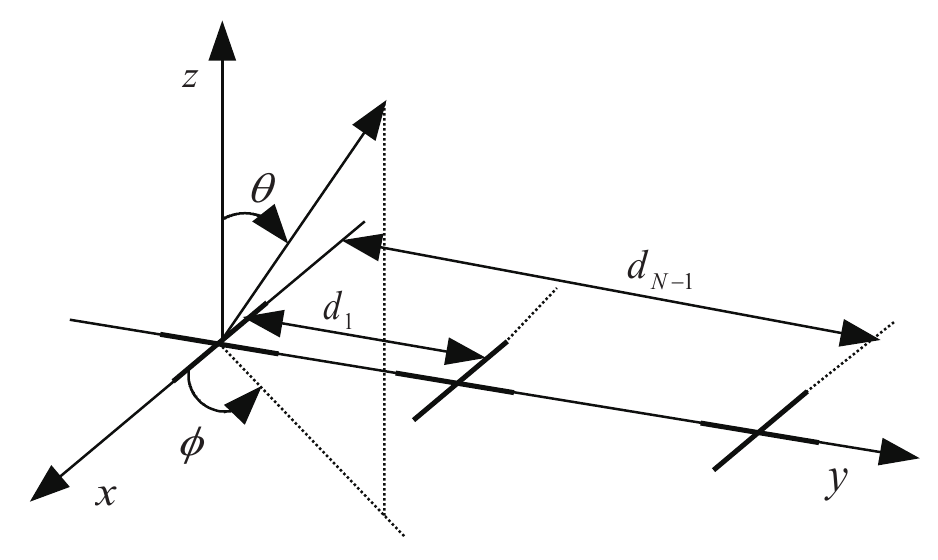}\\
  \caption{A crossed-dipole linear array.}\label{fig:crossed_dipole_array}
\end{figure}

A narrowband linear array for transmit beamforming based on $N$ crossed-dipole antennas is shown in Fig. \ref{fig:crossed_dipole_array}. For each antenna, there are two orthogonally orientated dipoles, and the one parallel to the x axis is connected to a complex valued weight coefficient represented by $w_{n,x}$, and the one parallel to the y axis is connected to $w_{n,y}$ for $n=0, \ldots, N-1$. The spacing from the first antenna to its subsequent antennas is represented by $d_{n}$ for $n=1, \ldots, N-1$. The elevation angle is denoted by $\theta\in [0,\pi]$ and azimuth angle is represented by $\phi\in [0,2\pi]$. The spatial steering vector of the array, a function of elevation angle $\theta$ and azimuth angle $\phi$, is given by
\begin{equation}
    \textbf{s}_s(\theta,\phi)=[1, e^{-j\omega d_1\sin\theta\sin\phi/c}, \ldots, e^{-j\omega d_{N-1}\sin\theta\sin\phi/c}]^{T},
\end{equation}
where $\{\cdot\}^{T}$ is the transpose operation, and $c$ is the speed of propagation. For a uniform linear array (ULA) with a half-wavelength spacing ($d_{n}-d_{n-1} = \lambda/2$), the spatial steering vector can be simplified to
\begin{equation}
    \textbf{s}_s(\theta,\phi)=[1, e^{-j\pi\sin\theta\sin\phi}, \ldots, e^{-j\pi(N-1)\sin\theta\sin\phi}]^{T}.
\end{equation}
Moreover, the spatial-polarisation coherent vector of the signal is given by~\cite{zhang14c,lan17a}
\begin{equation}
\begin{split}
\textbf{s}_{p}(\theta,\phi,\gamma,\eta) = &\left[
             \begin{array}{lcl}
             -\sin\phi\cos\gamma+\cos\phi\cos\theta\sin\gamma e^{j\eta} \\
             \cos\phi\cos\gamma+\sin\phi\cos\theta\sin\gamma e^{j\eta}
             \end{array}
        \right]\\
         = &\left[
             \begin{array}{lcl}
             s_{px}(\theta,\phi,\gamma,\eta) \\
             s_{py}(\theta,\phi,\gamma,\eta)
             \end{array}
        \right],
\end{split}
\end{equation}
where $\gamma\in[0,\pi/2]$ is the auxiliary polarisation angle and $\eta\in[-\pi,\pi)$ represents the polarisation phase difference.

For convenience, we split the array structure into two sub-arrays: one is parallel to the x-axis and the other to the y-axis. Then, the steering vectors of the two sub-arrays are given by
\begin{equation}
\begin{split}
\textbf{s}_{x}(\theta,\phi,\gamma,\eta) =&
                          s_{px}(\theta,\phi,\gamma,\eta)\textbf{s}_s(\theta,\phi),\\
\textbf{s}_{y}(\theta,\phi,\gamma,\eta) =&
                          s_{py}(\theta,\phi,\gamma,\eta)\textbf{s}_s(\theta,\phi).
\end{split}
\end{equation}
The beam response of the array can be given by~\cite{hawes15b}
\begin{equation}
\label{eq:response}
    p(\theta,\phi,\gamma,\eta)=\textbf{w}^{H}\textbf{s}(\theta,\phi,\gamma,\eta),
\end{equation}
where $\{\cdot\}^{H}$ represents the Hermitian transpose, $\textbf{s}(\theta,\phi,\gamma,\eta)$, a $2N\times 1$ vector representing the overall steering vector of the array, is given by
\begin{equation}
\begin{split}
\textbf{s}(\theta,\phi,\gamma,\eta) = &[\textbf{s}_{x}(\theta,\phi,\gamma,\eta),\textbf{s}_{y}(\theta,\phi,\gamma,\eta)]^T,\\
\textbf{s}_{x}(\theta,\phi,\gamma,\eta) = &[s_{0,x}(\theta,\phi,\gamma,\eta),\ldots,s_{N-1,x}(\theta,\phi,\gamma,\eta)]^T,\\
\textbf{s}_{y}(\theta,\phi,\gamma,\eta) = &[s_{0,y}(\theta,\phi,\gamma,\eta),\ldots,s_{N-1,y}(\theta,\phi,\gamma,\eta)]^T,\\
\end{split}
\end{equation}
and $\textbf{w}$ is the complex valued weight vector
\begin{equation}
    \textbf{w} = [w_{0,x}, \ldots, w_{N-1,x},w_{0,y}, \ldots, w_{N-1,y}]^{T}.
\end{equation}

\section{Directional Modulation Design}\label{sec:Directional Modulation Design}

In traditional crossed-dipole array design, the weight coefficients are designed for one single beam response and therefore can only cater for one transmitted DM signal. To increase the transmission capacity, in this section, we design a set of common weight coefficients for two signals ($S_1$ and $S_2$) with orthogonal polarisation states transmitted to the same direction simultaneously. Based on the parameter representations in Sec. \ref{sec:Review of Polarised Beamforming}, we have $\textbf{s}(\theta,\phi,\gamma_1,\eta_1)$, $\textbf{s}(\theta,\phi,\gamma_2,\eta_2)$, $p(\theta,\phi,\gamma_1,\eta_1)$, and $p(\theta,\phi,\gamma_2,\eta_2)$ to represent steering vectors for $S_1$ and $S_2$, and beam responses for $S_1$ and $S_2$, respectively.

For $M$-ary signaling, such as multiple phase shift keying (MPSK), each of $S_1$ and $S_2$ can create $M$ constellation points ($M$ symbols), leading to $M$ desired responses. Then, for $S_1$ and $S_2$ transmitted simultaneously, there are $M^2$ different symbols and $M^2$ sets of response pairs, represented by a $1\times 2$ vector $\textbf{p}_m(\theta,\phi,\gamma,\eta)$ for $m = 0,\ldots, M^2-1$, and each response $\textbf{p}_m(\theta,\phi,\gamma,\eta)$ corresponds to a weight vector $\textbf{w}_{m}=[w_{0,x,m}, \ldots, w_{N-1,x,m},w_{0,y,m}, \ldots, w_{N-1,y,m}]^T$. Here, we assume $r$ points are sampled in the mainlobe and $R-r$ points in the sidelobe range; then the responses for two signals in the mainlobe direction $\textbf{p}_{ML,m}$ and the sidelobe range $\textbf{p}_{SL,m}$ are given by
\begin{equation}
\begin{split}
\textbf{p}_{SL,m} = [&p_{m}(\theta_{0},\phi,\gamma_1,\eta_1), \ldots, p_{m}(\theta_{R-r-1},\phi,\gamma_1,\eta_1)\\
&p_{m}(\theta_{0},\phi,\gamma_2,\eta_2), \ldots, p_{m}(\theta_{R-r-1},\phi,\gamma_2,\eta_2)],\\
\textbf{p}_{ML,m} = [&p_{m}(\theta_{R-r},\phi,\gamma_1,\eta_1), \ldots, p_{m}(\theta_{R-1},\phi,\gamma_1,\eta_1)\\
&p_{m}(\theta_{R-r},\phi,\gamma_2,\eta_2), \ldots, p_{m}(\theta_{R-1},\phi,\gamma_2,\eta_2)].
 \end{split}
\end{equation}

Here we have implicitly assumed that all the responses are considered with a fixed $\phi$. Moreover, we put all the $2(R-r)$ steering vectors of both $S_1$ and $S_2$ at the sidelobe region into an $2N\times 2(R-r)$ matrix $\textbf{S}_{SL}$, and the $2r$ steering vectors at the mainlobe direction into an $2N\times 2r$ matrix, denoted by $\textbf{S}_{ML}$, where
\begin{equation}
\begin{split}
\textbf{S}_{SL} = [&\textbf{s}(\theta_{0},\phi,\gamma_1,\eta_1),\ldots,\textbf{s}(\theta_{R-r-1},\phi,\gamma_1,\eta_1)\\
&\textbf{s}(\theta_{0},\phi,\gamma_2,\eta_2),\ldots,\textbf{s}(\theta_{R-r-1},\phi,\gamma_2,\eta_2)],\\
\textbf{S}_{ML} = [&\textbf{s}(\theta_{R-r},\phi,\gamma_1,\eta_1),\ldots,\textbf{s}(\theta_{R-1},\phi,\gamma_1,\eta_1)\\
&\textbf{s}(\theta_{R-r},\phi,\gamma_2,\eta_2),\ldots,\textbf{s}(\theta_{R-1},\phi,\gamma_2,\eta_2)].
\end{split}
\end{equation}
Then, for the $m$-th symbol, its corresponding weight coefficients can be solved by
\begin{equation}
\label{eq:dm_with_a_given_array}
\begin{split}
    \min\quad&||\textbf{p}_{SL,m}-\textbf{w}_{m}^{H}\textbf{S}_{SL}||_{2}\\
    \text{subject to}\quad&\textbf{w}_{m}^{H}\textbf{S}_{ML} = \textbf{p}_{ML,m},
\end{split}
\end{equation}
where $||\cdot||_2$ denotes the $l_2$ norm. The objective function in (\ref{eq:dm_with_a_given_array}) ensures a minimum difference between the desired and designed responses for two signals in the sidelobe, and the constraint keeps a desired constellation value of two signals to the mainlobe. To ensure that the constellation is scrambled in the sidelobe regions, the phase of the desired response for two signals $\textbf{p}_{SL,m}$ at different sidelobe directions can be randomly generated.

The problem in \eqref{eq:dm_with_a_given_array} can be solved by the method of Lagrange multipliers, and the optimum value for the weight vector $\textbf{w}_{m}$ is given in \eqref{eq:w_long_equations}, where $\textbf{R} = \textbf{S}_{SL}\textbf{S}_{SL}^H$.
\begin{equation}
\begin{split}
\label{eq:w_long_equations}
\textbf{w}_{m} = &\textbf{R}^{-1}(\textbf{S}_{SL}\textbf{p}_{SL,m}^H-\textbf{S}_{ML}
((\textbf{S}^H_{ML}\textbf{R}^{-1}\textbf{S}_{ML})^{-1}\\
&(\textbf{S}^H_{ML}\textbf{R}^{-1}\textbf{S}_{SL}\textbf{p}_{SL,m}^H
-\textbf{p}^H_{ML,m}))).
\end{split}
\end{equation}
As these two signals $S_1$ and $S_2$ are modulated to generate a 4-D modulated signal $S_{Re}$ in the far-field, to separate the corresponding components from the received composite signal, at the receiver side, we need a crossed-dipole antenna array in the same way as given in~\cite{liu14a}.

\begin{figure}
  \centering
  \subfigure[]{
  \includegraphics[width = 0.45\textwidth]{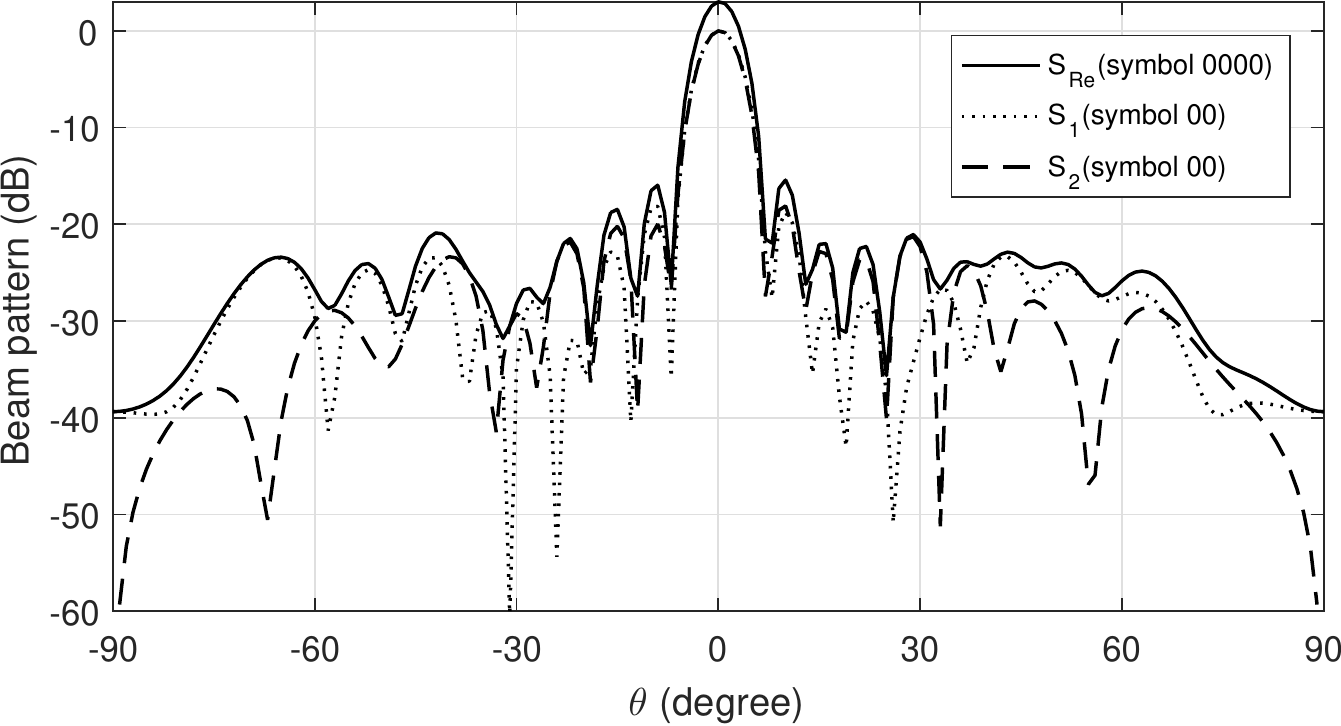}
  \label{fig:response_ula_h_v_channels1}}\\
  \subfigure[]{
  \includegraphics[width=0.45\textwidth]{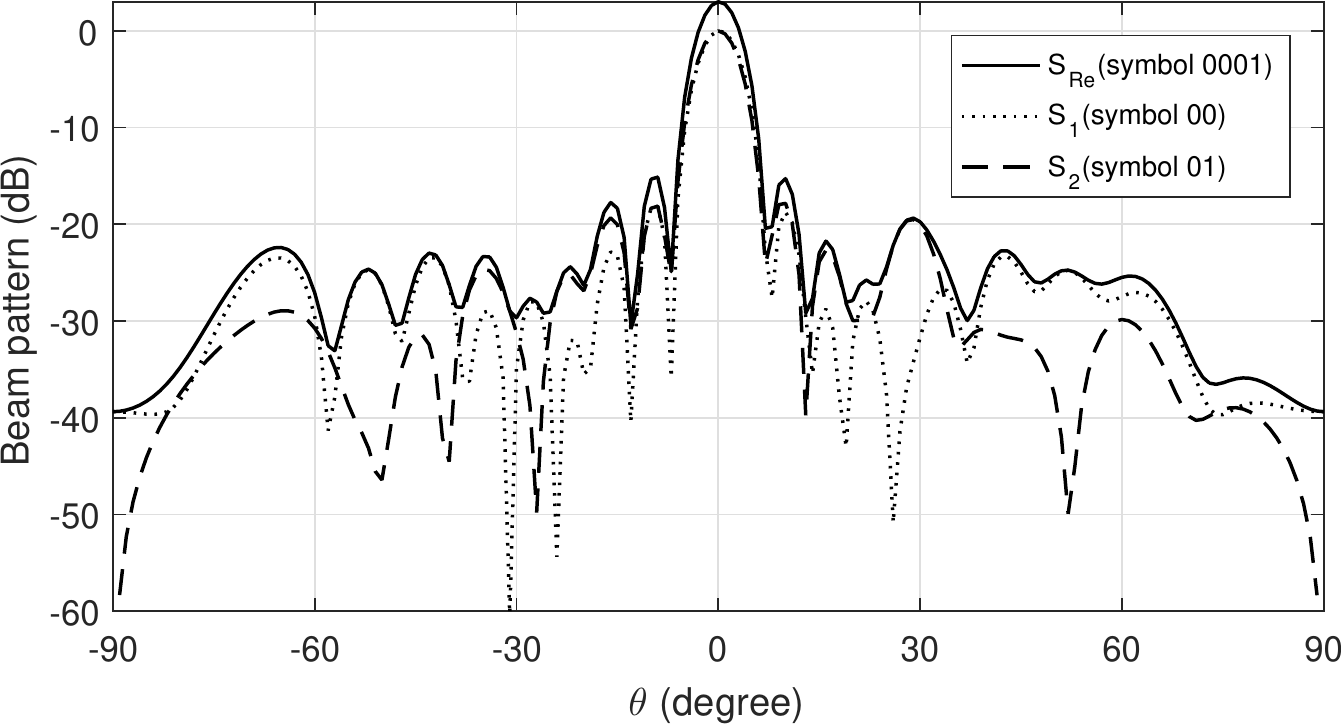}
  \label{fig:response_ula_h_v_channels2}}\\
  \subfigure[]{
  \includegraphics[width=0.45\textwidth]{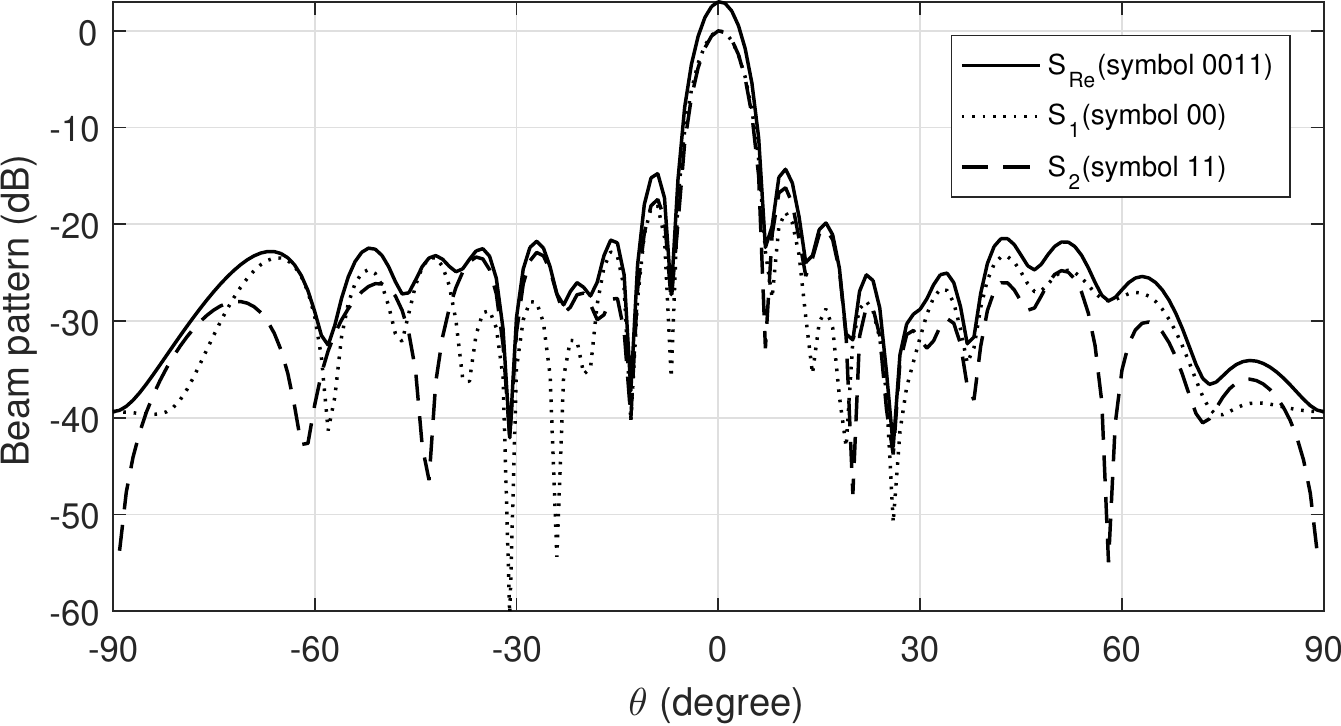}
  \label{fig:response_ula_h_v_channels3}}\\
  \subfigure[]{
  \includegraphics[width=0.45\textwidth]{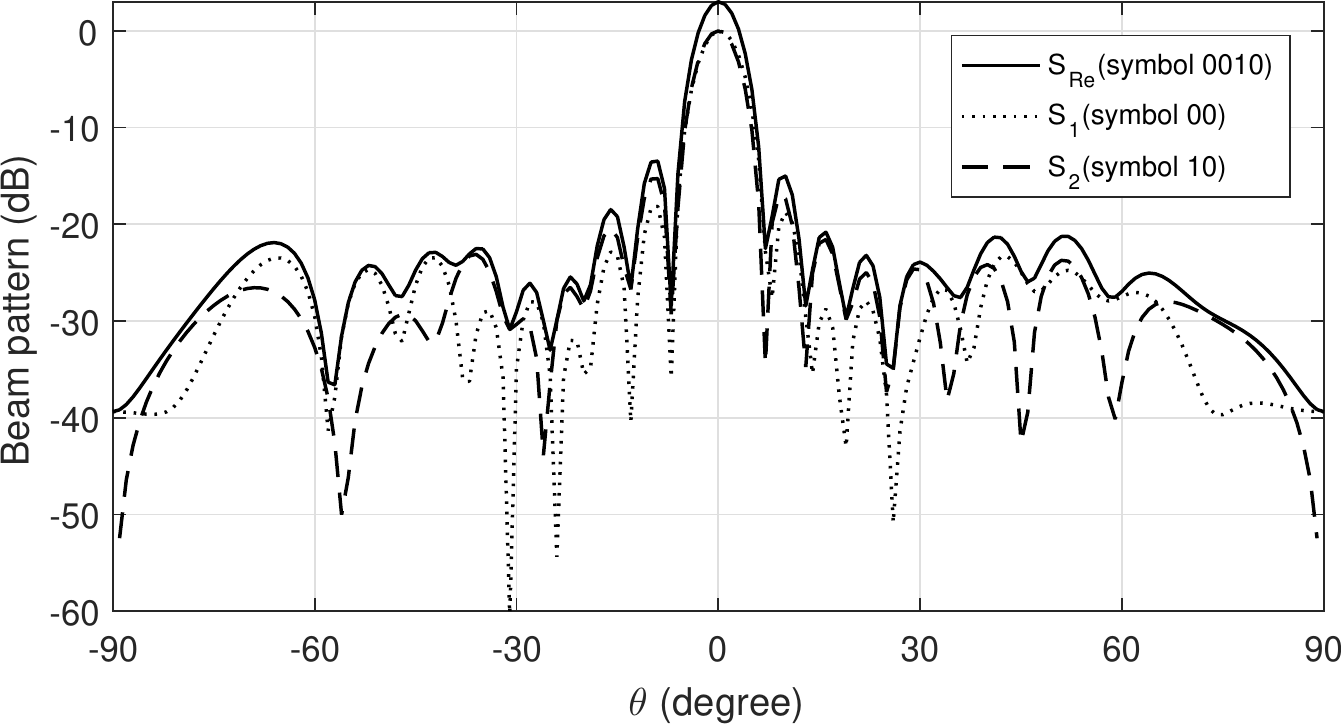}
  \label{fig:response_ula_h_v_channels4}}
  \caption{Resultant beam responses based on the design \eqref{eq:dm_with_a_given_array} for symbols (a) `00,00', (b)`00,01', (c)`00,11', (d)`00,10'.}
\end{figure}

\begin{figure}
  \centering
  \subfigure[]{
  \includegraphics[width = 0.45\textwidth]{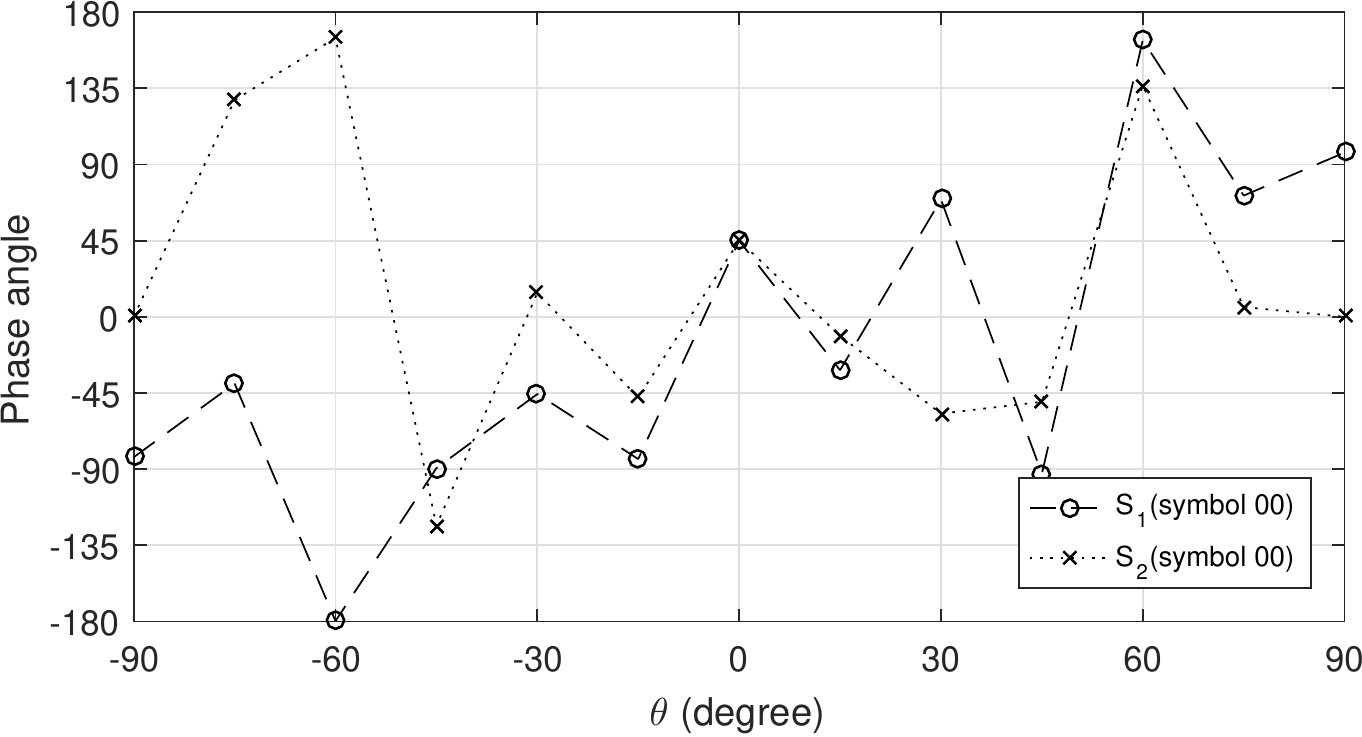}
  \label{fig:phase_ula_h_v_channels1}}\\
  \subfigure[]{
  \includegraphics[width=0.45\textwidth]{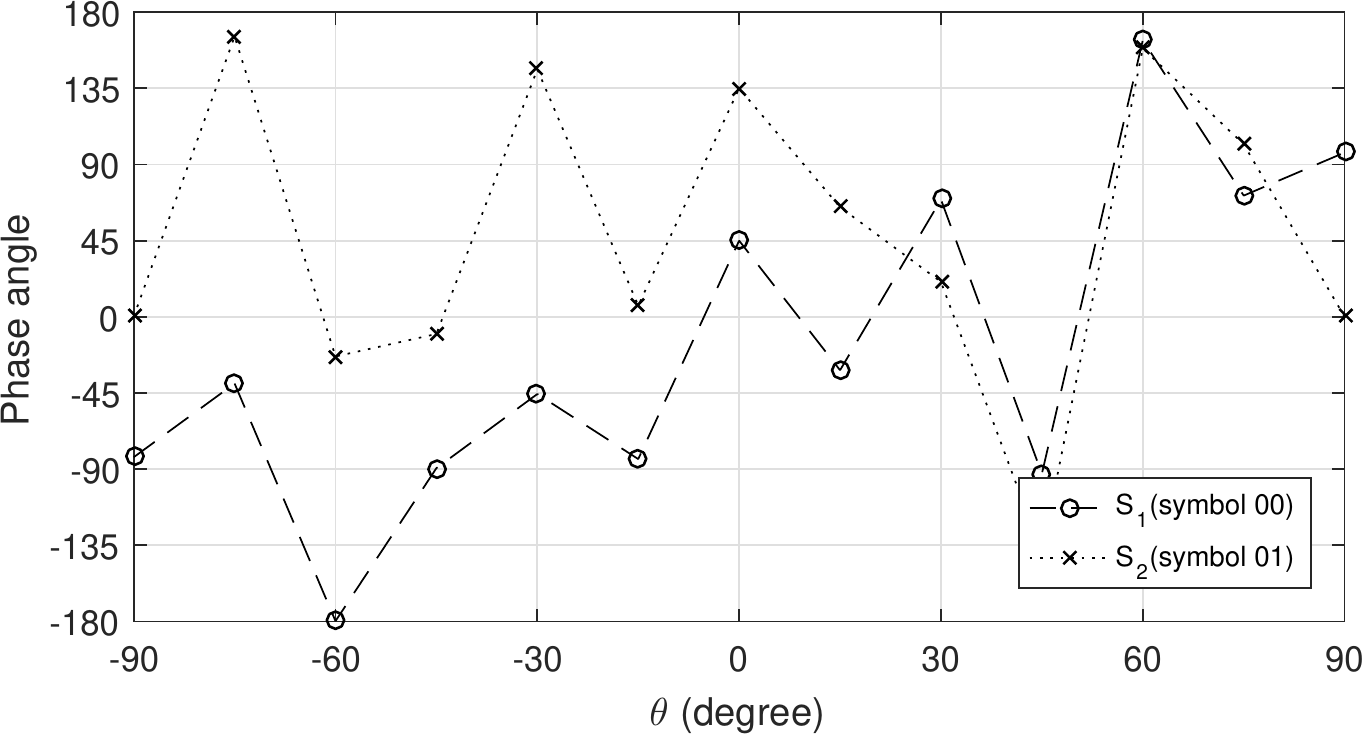}
  \label{fig:phase_ula_h_v_channels2}}\\
  \subfigure[]{
  \includegraphics[width=0.45\textwidth]{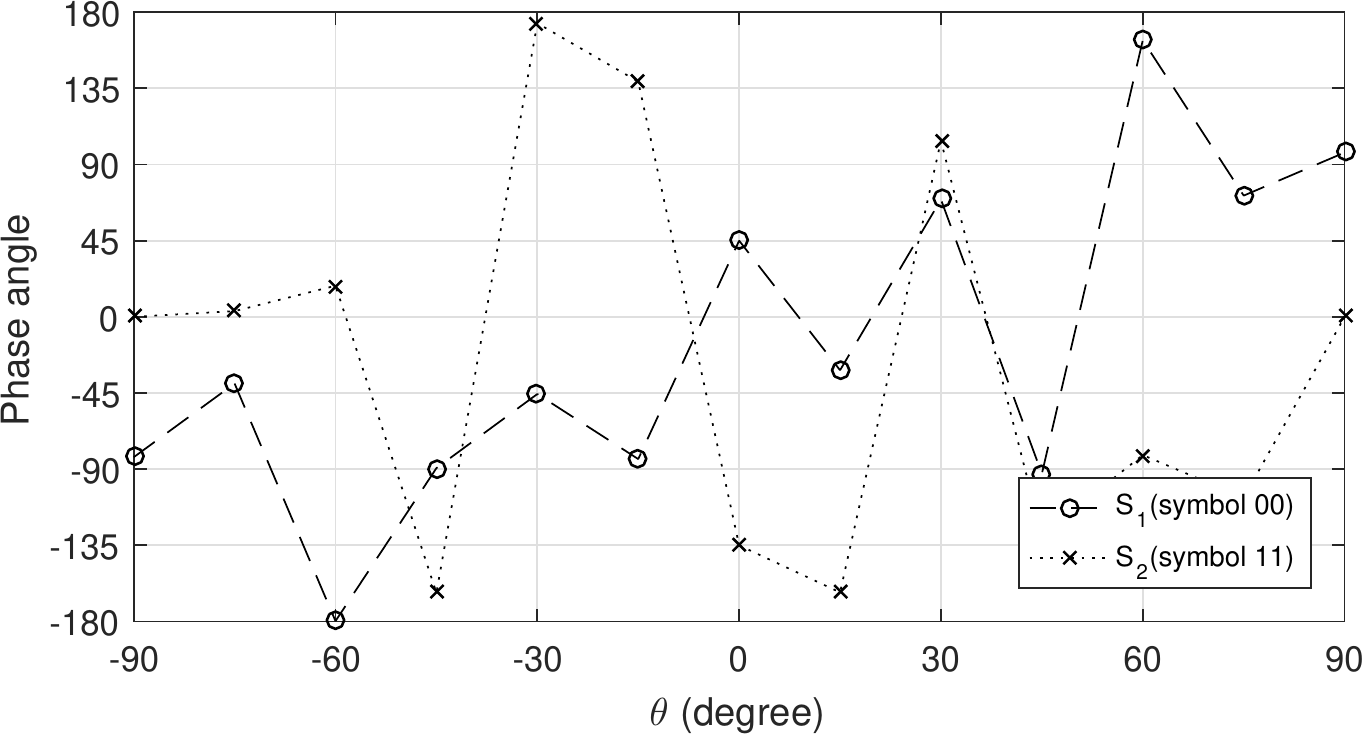}
  \label{fig:phase_ula_h_v_channels3}}\\
  \subfigure[]{
  \includegraphics[width=0.45\textwidth]{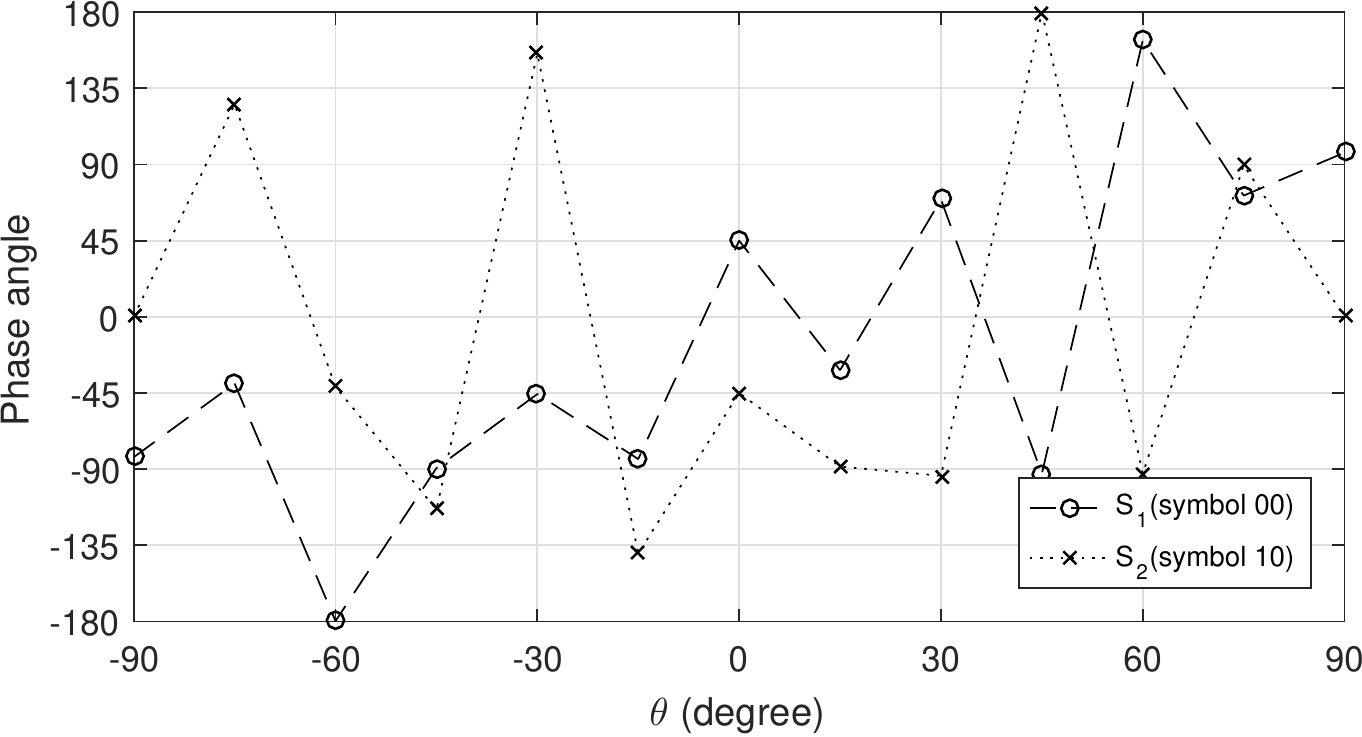}
  \label{fig:phase_ula_h_v_channels4}}
  \caption{Resultant phase responses based on the design \eqref{eq:dm_with_a_given_array} for symbols (a) `00,00', (b)`00,01', (c)`00,11', (d)`00,10'.}
\end{figure}

\section{Design examples}\label{sec:sim}

In this section, we provide design examples based on a $9\lambda$ aperture ULA with a half wavelength spacing between adjacent antennas to show the performance of the proposed formulations. The polarisation characteristics are defined by $(\gamma_1,\eta_1) = (0^\circ,0^\circ)$ for $S_1$ for a horizontal polarisation, and $(\gamma_2,\eta_2) = (90^\circ,0^\circ)$ for $S_2$ for a vertical polarisation. For each of $S_1$ and $S_2$, the desired response is a value of one (magnitude) with $90^\circ$ phase shift at the mainlobe (QPSK), i.e., symbols `00', `01', `11', `10' correspond to $45^{\circ}$, $135^{\circ}$, $-45^{\circ}$ and $-135^{\circ}$, respectively, and a value of $0.1$ (magnitude) with random phase shifts over the sidelobe regions. Therefore, for signals $S_1$ and $S_2$ transmitted simultaneously, we can construct $16$ different symbols. Moreover, without loss of generality, we assume the mainlobe direction is $\theta_{ML} = 0^{\circ}$ for $\phi = 90^{\circ}$ and the sidelobe regions are $\theta_{SL} \in [5^{\circ}, 90^{\circ}]$ for $\phi = \pm 90^{\circ}$, sampled every $1^{\circ}$, representing two signals are transmitted through the $y-z$ plane.

With the above parameters and design formulations in \eqref{eq:dm_with_a_given_array}, the composite beam patterns for symbols `00,00', `00,01', `00,11' and `00,10' are shown in Figs. \ref{fig:response_ula_h_v_channels1}, \ref{fig:response_ula_h_v_channels2}, \ref{fig:response_ula_h_v_channels3} and \ref{fig:response_ula_h_v_channels4}, where all main beams are exactly pointed to $0^\circ$ (3.01dB or $\sqrt{2}$ magnitude) with a reasonable sidelobe level, and its corresponding components for $S_1$ and $S_2$ coincide in the mainlobe direction with $0$dB mainlobe level, representing that a value of one for magnitude of desired signals $S_1$ and $S_2$ has been satisfied. The phase patterns for these symbols are displayed in Figs. \ref{fig:phase_ula_h_v_channels1}, \ref{fig:phase_ula_h_v_channels2}, \ref{fig:phase_ula_h_v_channels3}, \ref{fig:phase_ula_h_v_channels4}. We can see that the phases of $S_1$ and $S_2$ in the desired direction match the standard QPSK constellation, while for the rest of the $\theta$ angles, phases are random, demonstrating that directional modulation has been achieved effectively. The beam and phase patterns for other symbols are not shown as they have the same features as the aforementioned figures.

\section{Conclusions}\label{sec:con}
A crossed-dipole antenna array for directional modulation has been proposed for the first time, and a set of common weight coefficients has been designed for two signals $S_1$ and $S_2$ with orthogonal polarisation states transmitted to the same direction, based on a fixed given array geometry. As these two signals are modulated to generate a 4-D modulated signal $S_{Re}$ in the far-field, to receive and separate the corresponding components from the resultant $S_{Re}$, at the  receiver side, a crossed-dipole antenna array is needed. As shown in the provided design examples, for both $S_1$ and $S_2$, a mainlobe and standard phase pattern have been achieved in the desired direction with low sidelobe level and scrambled phases in other directions.

\renewcommand\refname{Reference}
\bibliographystyle{IEEEtran}
\bibliography{mybib}
\end{document}